\documentclass[prb,showpacs]{revtex4}
\usepackage{graphicx,psfrag,amssymb,amsmath}
\begin{document}
\title{Large-$N$ limit of a magnetic impurity in unconventional density waves}
\author{Bal\'azs D\'ora}
\affiliation{The Abdus Salam ICTP, Strada Costiera 11, I-34014, Trieste, Italy}

\date{\today}

\begin{abstract}
We investigate the effect of unconventional density wave (UDW) condensate on an Anderson impurity using large-$N$ technique at $T=0$.
In accordance with previous treatments of a Kondo impurity in pseudogap phases, we find that Kondo effect occurs only in a certain 
range of parameters. The f-electron density of states reflects the influence of UDW at low energies and around the maximum of the 
density wave gap. The static spin susceptibility diverges at the critical coupling, indicating the transition from strong to weak 
coupling. In the dynamic spin susceptibility an additional peak appears showing the presence the UDW gap. Predictions concerning 
non-linear density of states are made.
Our results apply to other unconventional condensates such as d-wave superconductors and d-density waves as 
well.
\end{abstract}


\maketitle

\section{Introduction}

Our understanding of the problem of dilute concentration of magnetic impurities in normal metals benefited a lot from exact 
solutions, and from reliable approximate methods\cite{hewson}. Among the latter,
the large-$N$ expansion seems to describe successfully the low temperature properties of matter, $N$ denotes the spin degeneracy, and 
$1/N$ can 
be used as an expansion parameter\cite{bickers,coleman,advphys}.
On the other hand, recently considerable attention has been focused on the behaviour of magnetic impurities in pseudogap phases, where 
the conduction electron density of states varies as a power-law of energy around the Fermi 
energy\cite{WF,borkowski,cassanello,zhang,zhu,polkov,buxton}. 
The interest on this subject 
mainly arises due to the behaviour of the pseudogap and superconducting phase of high $T_c$ superconductors. The interplay between 
unconventional condensates and quantum magnetic impurities can reveal the nature of the underlying phase. Using large-$N$ technique, 
several interesting results were found concerning the transition from magnetic to non-magnetic phases, the quasiparticle density of 
states etc. It was stated that particle-hole symmetry breaking is necessary to enter into the Kondo regime\cite{polkov}, which 
condition is 
naturally satisfied with an infinite-U Anderson impurity.

Experimentally, from scanning tunneling microscopy studies in 
Bi$_2$Sr$_2$CaCu$_2$O$_{8+\delta}$\cite{hudson,pan}, a strong variation of the electron density of states was found around the impurity 
site: 
beyond the superconducting coherence peaks, new structures were identified indicating the presence of impurity induces bound 
states.
Some of the experiments can be explained by pure potential scattering, but other features
 call for theoretical works to understand the effect of unconventional condensates on magnetic impurities\cite{stm}.

This is why we have chosen to study the interaction between unconventional density waves (UDW) and magnetic impurities. In UDW, the gap 
in the quasiparticle spectrum vanishes on certain subsets of the Fermi surface, and its average over the Fermi surface is 
zero\cite{nagycikk,nayak}. This 
causes the lack of periodic modulation of the spin and charge density. Such states have been proposed to describe the 
pseudogap phase of high $T_c$ superconductors\cite{nayak,moca,benfatto}, the low temperature phase of 
$\alpha$-(BEDT-TTF)$_2$KHg(SCN)$_4$\cite{rapid, tesla}, the antiferromagnetic phase in URu$_2$Si$_2$\cite{IO,roma} and other heavy 
fermion materials\cite{GG}, the 
charge density wave in 2H-TaSe$_2$\cite{castroneto} and the pseudogap phase in transition metal oxides\cite{3dflux}.
Due to the wavevector dependence of 
the gap, the transition to this 
phase is metal to metal instead of metal to insulator, as in conventional density waves (with constant energy gap).

The paper is organized as follows: in Sec. II, we study an Anderson impurity embedded in an unconventional density 
wave in the large-$N$ limit at $T=0$. As opposed to previous treatments\cite{zhang,zhu}, we allow for a macroscopic occupation of the 
$f$ state. From the 
saddle point equations, the phase diagram is determined, and the effect of magnetic 
field is discussed. In Sec. III, we turn to the investigation of the properties of the impurity. Its density of states displays the 
same power-law energy dependence at low energies as that of band electrons, enhances around $\pm\Delta$ and a Kondo peak shows up 
at positive energies. The presence of these three peaks is in accord with experimental findings in 
Bi$_2$Sr$_2$CaCu$_2$O$_{8+\delta}$\cite{hudson,pan}.
The peculiarities of the 
conduction electron transport lifetime are explored.
The static spin susceptibility signals the transition from the Kondo to the decoupled free moment regime. The dynamic spin 
susceptibility exhibits the usual Kondo peak, plus an additional peak coming from the divergent peak in the density of states of UDW.
Some generalizations to non-linear density of states are made.
In spite of the different topology of the Fermi surfaces, our results apply to other phases with power-law density of states like in 
d-wave superconductors\cite{szummad-wave} or in d-density 
waves\cite{sudip}.

\section{Phase diagram}

The Hamiltonian describing an infinite-U Anderson impurity interacting with an unconventional density wave is given by:
\begin{gather}
 H=\sum_{{\bf k},m}^\prime\left[\xi({\bf k})(a_{{\bf k},m}^{+}a_{{\bf
 k},m}-a_{{\bf k-Q},m}^{+}a_{{\bf
 k-Q},m})+\Delta({\bf k})(a_{{\bf k},m}^{+}a_{{\bf
 k-Q},m}+a_{{\bf k-Q},m}^{+}a_{{\bf  k},m})+\right.\nonumber\\
\left. \frac{V}{\sqrt N}\left((a^+_{{\bf k},m}+a^+_{{\bf k-Q},m})f_m b^+ +
f^+_m\left(a_{{\bf k},m}+a_{{\bf k-Q},m}\right)b\right)\right]+E\sum_m f^+_mf_m ,
\label{hamilton}
\end{gather}
where $a_{{\bf k},m}^{+}$ and $a_{{\bf k},m}$ are, respectively,
the creation and annihilation operators of an electron of momentum $\bf k$ and
spin $m$, $-s\leq m \leq s$, $N=2s+1$. Similarly, $f^+_m$ and $f_m$ creates and annihilates an electron on the localized $E$ level,
$b^+$ and $b$ are the slave boson operators, responsible for the hole states\cite{bickers,coleman}.
In a sum with prime, $k_x$ runs from $0$ 
to $2k_F$ ($k_F$ is the Fermi wavenumber), ${\bf 
Q}=(2k_F,\pi/b,\pi/c)$ is 
the best nesting vector. $\Delta({\bf k})=\Delta\sin(bk_y)$ is the unconventional density wave order parameter, and is taken to be 
real for 
simplicity. A gap with $\cos(bk_y)$, or $k_y\rightarrow k_z$ replacement would yield to the same results. 
Our system is based on an orthogonal lattice, with lattice constants $a,b,c$
toward direction $x,y,z$. The system is anisotropic, the quasi-one-dimensional
direction is the $x$ axis. The $a$ electron system possesses a density of states, which varies linearly with energy around the Fermi 
energy: $N(\omega)/N_0=|\omega|/\Delta$ for $|\omega|\ll\Delta$, $N_0$ is the normal state density of states per spin at the Fermi 
energy\cite{nagycikk}.
The linearized kinetic-energy spectrum is $\xi({\bf k})=v_F(k_x-k_F)$, and the dispersion in 
the other directions can be neglected for practical purposes.
By introducing the spinor
\begin{equation}
\Psi({\bf k},m,\tau)=
\left( \begin{array}{c}
         a_{{\bf k},m}(\tau) \\
         a_{{\bf k-Q},m}(\tau) 
         \end{array}
 \right),
\end{equation}
the single-particle thermal Green's function of the $a$ electrons without hybridization is obtained from Eq. (\ref{hamilton})
as
\begin{equation}
G({\bf k}, i\omega_n)=-\int_0^\beta d\tau\langle T_\tau \Psi({\bf k},m,\tau)
\Psi^+({\bf k},m,0)\rangle_He^{i\omega_n\tau}=\left[i\omega_n-\xi({\bf
k})\rho_3-\rho_1\Delta({\bf
k})\right]^{-1},
\end{equation}
where $\omega_n$ is the fermionic Matsubara frequency, $\rho_i$'s ($i=1,2,3$) are the usual Pauli
matrices
acting on
momentum space.

\begin{figure}[h!]
\psfrag{x}[t][b][1.2][0]{$E/\Delta$}
\psfrag{y}[b][t][1.2][0]{$b_0$}
{\includegraphics[width=7cm,height=7cm]{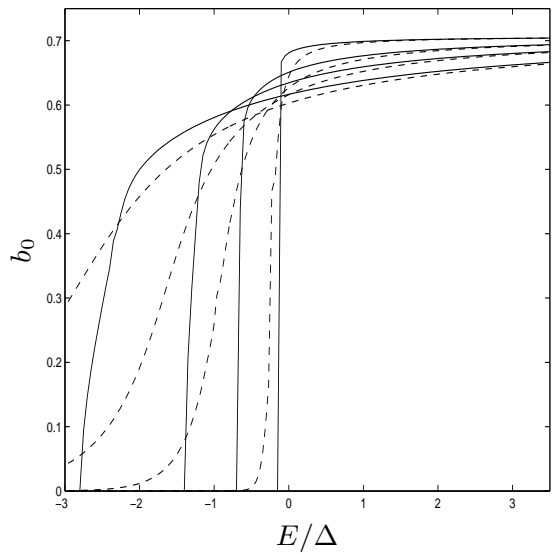}}
\caption{The $E$ dependence of the order parameter (the expectation value of the slave-boson operator) is plotted for $q=1/2$, $W=20\Delta$, 
$\Gamma/\Delta=0.1$, $0.5$, $1$ and $2$ from right to left (solid line), while the dashed line represents the behaviour of an 
Anderson impurity embedded in a normal metallic host.}
\label{bvse2}
\end{figure}

\begin{figure}[h!]
\psfrag{x}[t][b][1.2][0]{$E/\Delta$}
\psfrag{y}[b][t][1.2][0]{$b_0$}
{\includegraphics[width=7cm,height=7cm]{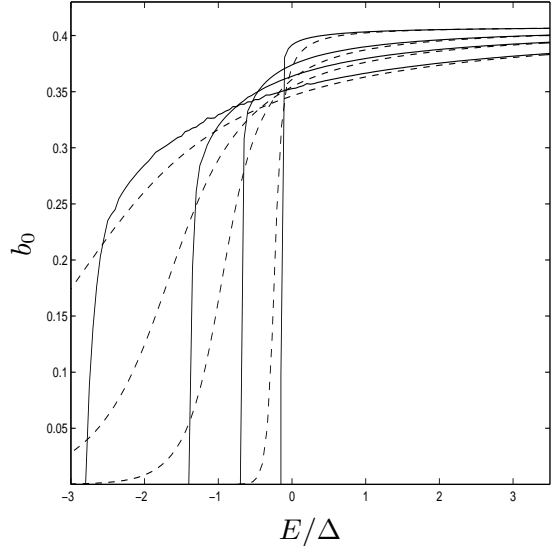}}
\caption{The $E$ dependence of the order parameter is shown for $q=1/6$, $W=20\Delta$,
$\Gamma/\Delta=0.1$, $0.5$, $1$ and $2$ from right to left (solid line), while the dashed line represents the behaviour of an
Anderson impurity embedded in a normal metallic host.}
\label{bvse6}
\end{figure}

The Hamiltonian should be restricted to the subspace
\begin{equation}
\sum_m f^+_mf_m+b^+b=Q
\end{equation}
where in the true large-$N$ limit $Q$ grows extensively with $N$, i.e. $Q=Nq$, $0<q<1$. In the original formulation of the model, $Q=1$, 
but in order to preserve
a macroscopic occupation of the $f$ state, the above generalization was found to be useful\cite{coleman}. It has been argued that the 
properties of 
the spin-$\frac 12$ Anderson model are best represented by $q=1/2$ in the large-$N$ limit\cite{bickers}.
Within the mean-field approximation, the slave-boson operators are replaced by their expectation value, $b_0=\langle b 
\rangle/\sqrt{N}$, and the constraint is 
satisfied by introducing a Lagrange multiplier $\lambda$:

\begin{gather}
 H=\sum_{{\bf k},m}^\prime\left[\xi({\bf k})(a_{{\bf k},m}^{+}a_{{\bf
 k},m}-a_{{\bf k-Q},m}^{+}a_{{\bf
 k-Q},m})+\Delta({\bf k})(a_{{\bf k},m}^{+}a_{{\bf
 k-Q},m}+a_{{\bf k-Q},m}^{+}a_{{\bf k},m})+\right.\nonumber\\
\left. Vb_0\left((a^+_{{\bf k},m}+a^+_{{\bf k-Q},m})f_m+
f^+_m\left(a_{{\bf k},m}+a_{{\bf k-Q},m}\right)\right)\right]+(E+\lambda)\sum_m f^+_mf_m+N\lambda(b_0^2-q) ,
\label{mfh}
\end{gather}
It has been shown\cite{coleman} that the slave boson mean-field approximation produces the correct low energy physics of the 
conventional Anderson 
impurity model in the entire parameter range at $T=0$. 
The value of $\lambda$ and $b_0$ is determined by minimizing the free energy of the system with respect to them\cite{zhang,zhu}. As a 
result, the 
saddle point equations at $T=0$ are given by:
\begin{gather}
b_0^2=q-\frac 12+\frac 1\pi \int_0^\infty \frac{E+\lambda}{x^2(1+\tilde\Gamma \alpha(x))^2+(E+\lambda)^2}dx,\label{spe1}\\
b_0\lambda=\frac {b_0}{\pi} \int_0^W\frac{x^2(1+\tilde\Gamma \alpha(x))\alpha(x)\Gamma}{x^2(1+\tilde\Gamma 
\alpha(x))^2+(E+\lambda)^2}dx,\label{spe2}
\end{gather}
where $\alpha(x)=2 K(1/\sqrt{1+(x/\Delta)^2})/\sqrt{\Delta^2+x^2}\pi$ for $x<W$, $0$ otherwise, $K(z)$ is the complete elliptic 
integral of the first 
kind. The case of a magnetic impurity embedded in a normal metallic host can be studied with $\alpha(x)=1/|x|$. $W=v_Fk_F$ is half of 
the 
bandwidth, $\Gamma=V^2\pi N_0$, 
$\tilde\Gamma=\Gamma b_0^2$. The critical value of $E$ is obtained as
\begin{equation}
E_c=-\frac{\Gamma}{\pi}\ln\left(\frac{4 W}{\Delta}\right)
\end{equation}
in the weak coupling limit ($W\gg \Delta$). Below this value, only the trivial solution of the saddle point equations exist, namely 
$b_0=0$ and $\lambda=-E$, and no Kondo effect occurs. To study the more general case, when the density of states varies as 
a power-law 
of energy, one has to assume $\Delta({\bf k})=\Delta|\sin(bk_y)|^r\textmd{sign}(\sin(bk_y))$, $0<r<\infty$. The sign function assures 
the vanishing average of the gap over the Fermi surface, and the $r$ exponent results in a density of states as 
$N(\omega)/N_0=|\omega/\Delta|^{1/r}\Gamma(1/2r)/\sqrt \pi r\Gamma(1+r/2r)$ for $|\omega/\Delta|^{1/r}\ll 1$, and $\Gamma(x)$ is 
the complete 
Gamma 
function. With this, the above condition is modified as 
\begin{equation}
E_c=-\frac{\Gamma}{\pi}\ln\left(\frac{2^{r+1}W}{\Delta}\right)
\end{equation}
The higher the value of $r$, the lower the critical $f$ level energy, and by letting $r\rightarrow\infty$, the case of a normal metal 
can 
be reached with a constant density of states, when $E_c\rightarrow-\infty$, which means, that Kondo effect is always present at $T=0$.
The saddle point equations have been solved numerically, and the results are shown in Figs. \ref{bvse2} and \ref{bvse6} for $q=1/2$ and 
$q=1/6$. For comparison, we also show the results
assuming a normal metallic host. As $\Gamma$ decreases, $b_0$ increases much sharper, and reaches its maximum value $\sqrt q$ 
rapidly. In spite of the different constraints ($q=1/2$ and $1/6$), the similarity between the figures is striking, indicating that the 
results hardly depend on the filling factors. As to $\lambda$, it equals to $|E|$ below its critical value, and approaches $0$ 
as 
the $f$ level energy is further increased.

The effect of magnetic field can readily be incorporated into the theory by adding the Zeeman term to the Hamiltonian:
\begin{equation}
H_{Zeeman}=-h\sum_m t_mf^+_mf_m,
\end{equation}
where the appropriate choice of $t_m$ can represent the two extreme limits\cite{withoff}: (1) the straightforward generalization of the 
SU(2) model, 
where $t_m=m/N$ and the magnetic field completely lifts the impurity degeneracy, or (2), when $t_m=\textmd{sign}(m)$, where two 
$N/2$-fold degenerate levels are introduced by the field.
For large fields, the critical $f$ level energy is obtained as
\begin{gather}
E_c\approx E_{c0}+|h|\left\{\begin{array}{cc}
         \left(\dfrac 12 -q\right) & \textmd{for case (1)} \\
         \textmd{sign}\left(\dfrac 12-q\right) & \textmd{for case (2)}
         \end{array}\right.,
\end{gather}
where $E_{c0}$ is the critical $f$ level energy without magnetic field.
For $q\leq 1/2$, which is thought to be the physically motivated case, an applied magnetic field enhances the critical $E$, hence 
destroys the Kondo region. This follows naturally from the fact, that we have $N/2$ states below $E$ generated by the magnetic field, 
so the highest occupied state at $T=0$ is below the actual $E$.
On the other hand, for $q\geq 1/2$, by the use a magnetic field, 
the system can be driven back into the 
strong coupling regime, but the physical realization of such a situation (i.e. $q\geq 1/2$) seems to be doubtful. Here, the highest 
occupied state generated by the magnetic field is above $E$, which makes the hybridization possible again.

\section{Density of states, spin susceptibility}

\begin{figure}[h!]
\psfrag{x}[t][b][1.2][0]{$\omega/\Delta$}
\psfrag{y}[b][t][1.2][0]{$\rho_f(\omega)\Delta$}
{\includegraphics[width=7cm,height=7cm]{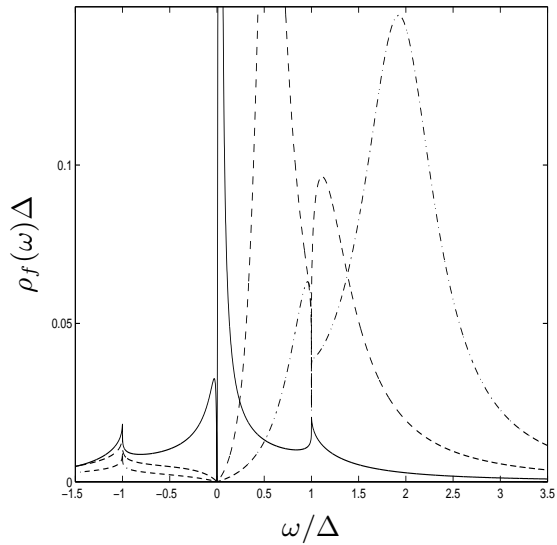}}
\caption{The $f$ electron density of states is shown for $q=1/4$ for $E$ slightly above $E_c$ (solid line), $E=0$ (dashed 
line) and $E=|E_c|$ (dashed-dotted line).}
\label{dosf}
\end{figure}

The Green's function of the $f$ electrons reads as
\begin{equation}
G_f(i\omega_n)=\frac{1}{i\omega_n-E-\lambda+i\omega_n\tilde\Gamma \alpha(\omega_n)},
\end{equation}
where $\alpha(x)$ was defined below Eq. \ref{spe2}.
The self energy of the $f$ electrons along the real frequency axis is obtained as
\begin{equation}
\Sigma_f(\omega+i\delta)=-\frac{2\tilde\Gamma}{\pi}\left[\Theta(\Delta-|\omega|)\left(\frac\omega\Delta K\left(\sqrt{1-
\left(\frac\omega\Delta\right)^2}\right)+i\left|\frac\omega\Delta\right| 
K\left(\frac\omega\Delta\right)\right)+\Theta(|\omega|-\Delta)iK\left(\frac\Delta\omega\right)\right],
\end{equation}
which exhibits the marginal Fermi liquid behaviour discovered by Zhang et al.\cite{zhang}: for small frequencies, the self energy 
varies 
as
$\Sigma_f(\omega\ll\Delta)=-\tilde\Gamma(2\omega\ln(4\Delta/|\omega|)/\pi+i|\omega|)/\Delta$. 
For $r\neq 1$ exponent, Re$\Sigma_f\sim \textmd{sign}(\omega)|\omega|^{\textmd{min}(1,1/r)}$, Im$\Sigma_f\sim |\omega|^{1/r}$. In 
general, the imaginary part of 
the self 
energy is directly proportional to the $a$ electron density of states.
From the self energy, the real $f$ electron density of states, which is the imaginary part of the Fourier transform of $-\langle 
\textmd{T}_\tau b^+(\tau)f_m(\tau)f^+_m(0)b(0)\rangle$\cite{coleman}, reads as
\begin{equation}
\rho_f(\omega)=-\frac{b_0^2}{\pi}\frac{\textmd{Im}\Sigma_f(\omega)}{(\omega-E-\lambda-\textmd{Re}
\Sigma_f(\omega))^2+(\textmd{Im}\Sigma_f(\omega))^2},
\label{fdos}
\end{equation}
which has the same low energy behaviour as the $a$ electron density of states, namely around the Fermi energy $\rho_f(\omega)\sim 
|\omega/\Delta|^{1/r}$. Also around $\pm\Delta$, it is expected to increase sharply due to the divergence of the UDW density of states, 
but the divergent peak is suppressed since the 
denominator in Eq. \ref{fdos} also diverges, as can be seen in Fig. \ref{dosf}. In fact, $\rho_f(\pm\Delta)=0$, which is invisible 
in Fig. \ref{dosf} due to its scale. The same phenomenon was discussed in charge density waves in the presence of a
non-magnetic impurity\cite{tutto2}. Above 
$E_c$, the large, asymmetric Kondo peak appears, and moves to higher frequencies, even above $\Delta$, as $E$ is 
further increased. 
%These structures in the density of states are similar to those found in 
%Bi$_2$Sr$_2$CaCu$_2$O$_{8+\delta}$\cite{hudson,pan}. Similar curves were obtained in Ref. \onlinecite{zhang,zhu}. 

\begin{figure}[h!]
\psfrag{x}[t][b][1.2][0]{$E/\Delta$}
\psfrag{y}[b][t][1.2][0]{$\Omega_K/\Delta$}
{\includegraphics[width=7cm,height=7cm]{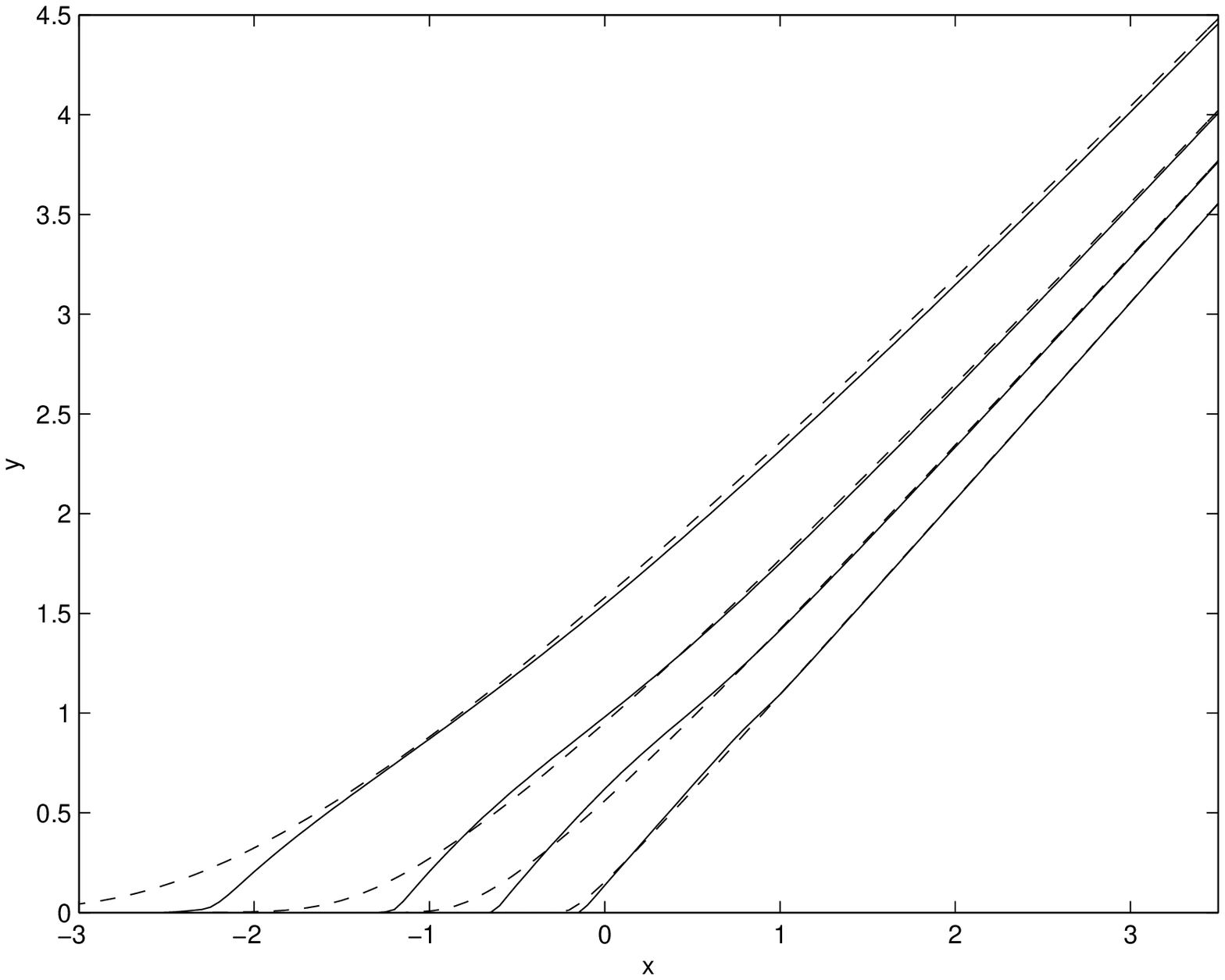}}
\psfrag{x}[t][b][1][0]{$E/\Delta$}
\psfrag{y}[b][t][1][0]{$\gamma/\Gamma$}

\vspace*{-6.8cm}\hspace*{-2cm}
\includegraphics[width=3.3cm,height=3.2cm]{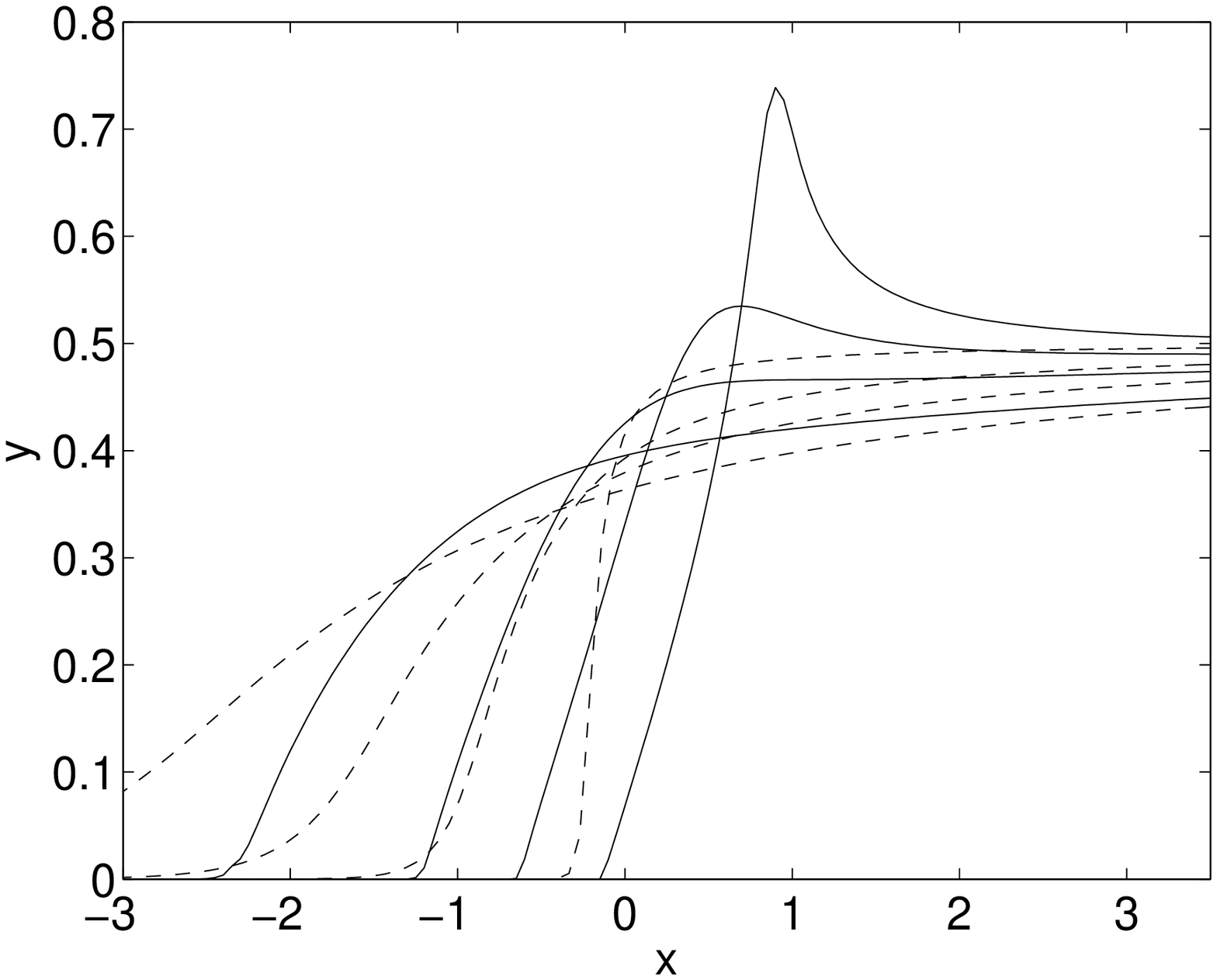}

\vspace*{3.5cm}
\caption{The Kondo scale, $\Omega_K$ is plotted for $q=1/2$, $\Gamma/\Delta=0.1$, $0.5$, $1$ and $2$ from right to left (solid line). 
The dashed 
line represents the behaviour of a normal metallic host. Note that $\Omega_K$ never vanishes in this case.
The inset shows the lifetime broadening, $\gamma$, together with normal host results.}
\label{kondotemp}
\end{figure}

The $T$-matrix of the conduction ($a$) electrons is obtained as
\begin{equation}
T(\omega+i\delta)=\frac{V^2 b_0^2}{\omega-E-\lambda-\Sigma_f(\omega)},
\label{tmatrix}
\end{equation} 
whose poles determine the impurity bound states or resonances\cite{moca}. By closely examining the denominator of Eq. \ref{tmatrix},
we find resonances at the $\Omega_K-i\gamma$, $\Omega_K$ can be identified with the Kondo temperature\cite{bickers,stm}, as was done in 
similar treatments of the large-$N$ limit of the 
Kondo model\cite{cassanello,polkov,zhang}, and $\gamma$ is the lifetime broadening of the $f$ level\cite{advphys}, and is shown in 
Fig. \ref{kondotemp}. The Kondo temperature increases almost linearly with $E$, $\gamma$ reaches its maximum around $E=\Delta$. For 
large $E$, the resonance appears at $E-iq\Gamma$, identically to the case of a normal conduction 
band\cite{advphys}. Alternatively, one can define the Kondo energy scale\cite{bickers,advphys} as $\sqrt{\Omega_K^2+\gamma^2}$, but it 
hardly differs from 
$\Omega_K$.

\begin{figure}[h!]
\psfrag{x}[t][b][1.2][0]{$\omega/\Delta$}
\psfrag{y}[b][t][1.2][0]{$\rho_a(\omega)/N_0$}
{\includegraphics[width=7cm,height=7cm]{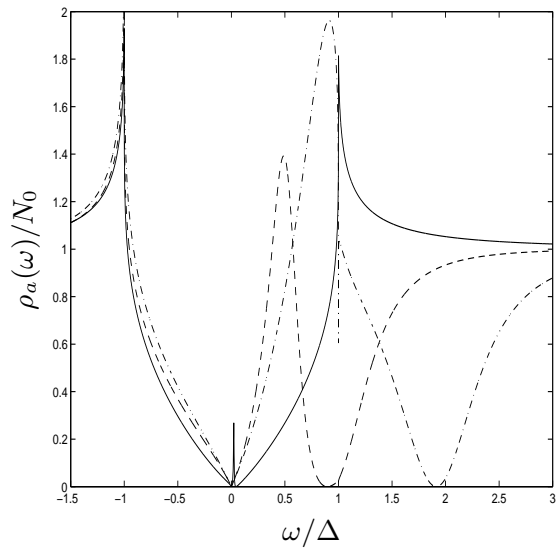}}
\caption{The local density of states (at the impurity site) of UDW is plotted for $q=1/4$ for $E$ slightly above $E_c$ 
(solid line), $E=0$ (dashed
line) and $E=|E_c|$ (dashed-dotted line).}
\label{dosa}
\end{figure}

The imaginary part of the $T$-matrix, which gives the inverse transport lifetime of the conduction electrons, is 
proportional to the real $f$ electron density of states, hence for small 
energies $\tau\sim |\omega|^{-1}$. Its detection is a very subtle issue, since the number of states having this behaviour is 
$N(\omega)\sim|\omega|$, hence $N(\omega)\tau\sim const$. The same cancellation is found for the $r\neq 1$ case.  Also $\tau$ is 
expected to be enhanced around 
$\pm\Delta$ as $\ln(2\sqrt 2/\sqrt{|1-|\omega/\Delta||})$. The number of possible states diverges in the same manner as $\tau$, 
resulting in $N(\omega)\tau\sim \ln^2\sqrt{|1-|\omega/\Delta||}$. For non-linear density of states, similarly to the case of 
$r=1$, both $\tau$ and $N(\omega)$ diverges around the gap maximum, which gives $N(\omega)\tau\sim\ln^2|1-|\omega/\Delta||^{1/2\sqrt 
r}$. This behaviour could be checked by experiments, although its detection is not an easy task, because such 
excitation are suppressed at low temperatures. 
With the help of the $T$-matrix, the local UDW density of states can be determined at the impurity site:
\begin{equation}
\rho_{a}(\omega)=\frac{N_0\rho_f(\omega)(\omega-E-\lambda)^2\pi}{\tilde\Gamma b_0^2},
\end{equation}
and is shown in Fig. \ref{dosa}. Slightly above $E_c$, a sharp resonance at $\Omega_K$ modifies the density 
of states. By increasing $E$, the Kondo peak is broadened, and the new zero at $\omega=E+\lambda$ can completely suppress the peak at 
$\omega=\Delta$.

Using Eq. \ref{fdos}, the low temperature specific heat is obtained as
\begin{equation}
C(T\rightarrow 0)=\frac{18\zeta(3)\tilde\Gamma}{\Delta\pi(E+\lambda)^2}T^2,
\end{equation}
which has the same temperature dependence as that of the density wave. For a non-linear density of states, the specific heat varies as
$C(T)\sim T^{(1+r)/r}$.
Also from the behaviour of $\rho_f(\omega)$, the low temperature behaviour of the resistivity can be calculated\cite{colemanaip}, 
and is 
obtained as 
$R(T)-R(0)\sim -T^2\ln^2(T)$. For general $r\neq 1$ exponents, it decreases as $-T^{\textmd{min}(2,2/r)}$.

\begin{figure}[h!]
\psfrag{x}[t][b][1.2][0]{$E/\Delta$}
\psfrag{y}[b][t][1.2][0]{$\chi(0)\Delta/\mu_{eff}^2$}
{\includegraphics[width=7cm,height=7cm]{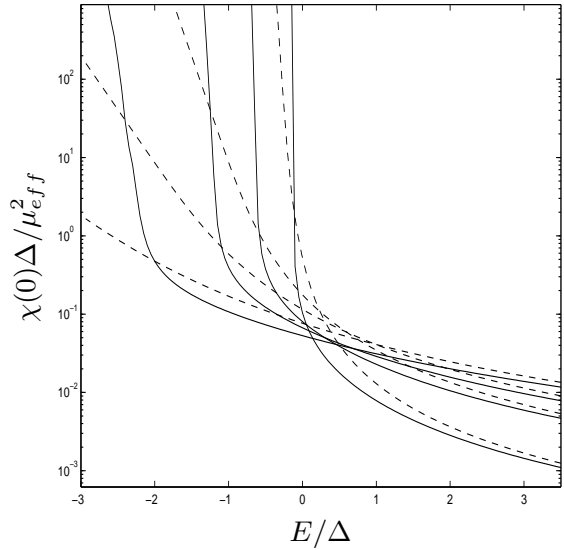}}
\caption{The $E$ dependence of static spin susceptibility is shown on a semilogarithmic scale for $q=1/2$, $W=20\Delta$,
$\Gamma/\Delta=0.1$, $0.5$, $1$ and $2$ from right to left (solid line), while the dashed line accounts for the susceptibility of an
Anderson impurity embedded in a normal metallic host.}
\label{suscvse2}
\end{figure}

The impurity spin susceptibility is calculated from
\begin{equation}
\chi(i\omega_m)=-\mu_{eff}^2T\sum_n G_f(i\omega_n+i\omega_m)G_f(i\omega_n),
\end{equation}
whose static limit at $T=0$ reads
as
\begin{equation}
\chi(0)=\mu_{eff}^2\left(\dfrac{b_0^2-q+\frac 12}{E+\lambda}-\frac 2\pi 
\int_0^\infty\frac{(E+\lambda)^2}{(x^2(1+\tilde\Gamma\alpha(x))^2+(E+\lambda)^2)^2}dx\right),
\end{equation}
which is shown in Fig. \ref{suscvse2} together with the corresponding results in a normal metallic host. The susceptibility diverges
at $E_c$, indicating the transition from the mixed valence to the decoupled free spin region, since in the latter  region, it behaves
as $\chi(T)=\mu_{eff}^2 q(1-q)/T$. 

\begin{figure}[h!]
\psfrag{x}[t][b][1.2][0]{$\omega/\Delta$}
\psfrag{y}[b][t][1.2][0]{Im$\chi(\omega)\Delta/\mu_{eff}^2$}
{\includegraphics[width=7cm,height=7cm]{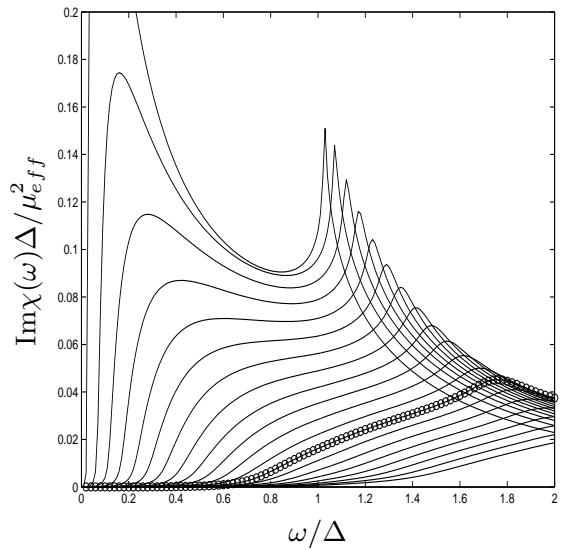}}
\caption{The evolution of the dynamic spin susceptibility for various values of $E$ (from $E=-1.25\Delta$ by $0.1\Delta$ steps from 
left to right) is shown 
for $q=1/6$, $W=20\Delta$,
$\Gamma/\Delta=1$. Two distinct peaks show up above $E_c$, and merge as one passes into the Kondo regime with increasing $E$. The 
circles denote $E\approx 0$.}
\label{dynsusc}
\end{figure}

The imaginary part of the impurity spin susceptibility should be readily accessible by neutron scattering experiments, and is 
evaluated at $T=0$ as
\begin{equation}
\textmd{Im}\chi(\omega)=\frac{\mu_{eff}^2\pi}{b_0^4}\int_0^\omega\rho_f(x)\rho_f(x-\omega)dx.
\end{equation}
For small frequencies, it behaves as Im$\chi(\omega)=\tilde\Gamma^2\omega^3/6\pi(E+\lambda)^4\Delta^2$, while for non-linear density 
of states, it increases as $\omega^{1+2/r}$. Beyond the usual Kondo peak occurring around $\Omega_K$, an additional sharp peak 
shows up at $\Delta+\Omega_K$ corresponding to excitations around the UDW gap. As $E$ increases, the latter becomes dominant, but 
its sharpness is smeared, as can be seen in Fig. \ref{dynsusc}. In a pure UDW, only one single peak is expected in Im$\chi(\omega)$ 
located at $2\Delta$, so the presence or absence of the extra two peaks can identify the presence or absence of magnetic impurities.

\section{Conclusion}

In summary we have studied the screening of a magnetic impurity in unconventional density waves. We considered an infinite-U 
Anderson impurity in the large-$N$ limit, when the occupation of the $f$ level increases extensively with $N$.
Kondo effect only occurs, if the $f$ level energy or the hybridization matrix element exceeds a certain value, and the result are 
almost independent of the chosen value of the $f$ level occupation.
The impurity density of states exhibits the UDW coherence peaks at $\pm\Delta$, and the usual Kondo peak. The conduction electron
density of states at the impurity site displays similar behaviour. The transport lifetime of the conduction electrons is determined 
from the $T$-matrix, and is found to diverge at $\omega=0$, and $\pm\Delta$. The former is suppressed due to the small number of states 
possessing this kind of behaviour.
The static impurity spin susceptibility diverges at the critical $f$ level energy, where the transition from Kondo to 
weak-coupling takes place. The dynamic spin susceptibility is enhanced at the Kondo energy, but an additional sharp resonance is 
noticed belonging to excitations from the gap maximum of UDW states to the Kondo peak. These features can help to identify the magnetic 
or non-magnetic nature of impurities in pseudogap phases.

%\begin{equation}
%\textmd{Im}\chi(\omega)=\frac{\mu_{eff}^2}{\pi}\int_{0}^\omega\textmd{Im}\left(\frac{1}{x-E-\lambda-\Sigma_f(x)}\right)
%\textmd{Im}\left(\frac{1}{x-\omega-E-\lambda-\Sigma_f(x-\omega)}\right)dx.
%\end{equation}

\begin{acknowledgments}
We are indebted to Gergely Zar\'and, Piers Coleman and Boris Narozhny for useful discussions.
\end{acknowledgments}
\bibliographystyle{apsrev}
\bibliography{anderson}
\end{document}